\documentclass[DIV=calc, paper=a4, fontsize=11pt, twocolumn]{scrartcl}	 

\usepackage{lipsum} 
\usepackage[english]{babel} 
\usepackage[protrusion=true,expansion=true]{microtype} 
\usepackage{amsmath,amsfonts,amsthm} 
\usepackage[svgnames]{xcolor} 
\usepackage[format=hang,width=.95\textwidth,justification=centering,small,labelfont=bf,textfont=it]{caption} 
\usepackage{booktabs} 
\usepackage{fix-cm}	 

\usepackage{sectsty} 
\allsectionsfont{\usefont{OT1}{phv}{b}{n}} 

\usepackage{fancyhdr} 
\usepackage{lastpage} 
\usepackage{natbib}
\usepackage{graphicx}
\usepackage{lettrine} 
\newcommand{\initial}[1]{ 
\lettrine[lines=3,lhang=0.3,nindent=0em]{
\color{DarkGoldenrod}
{\textsf{#1}}}{}}

\usepackage{titling} 

\newcommand{\HorRule}{\color{DarkGoldenrod} \rule{\linewidth}{1pt}} 

\pretitle{\vspace{-30pt} \begin{flushleft} \HorRule \fontsize{50}{50} \usefont{OT1}{phv}{b}{n} \color{DarkRed} \selectfont} 

\title{M\'{a}s all\'{a} del GWAS: alternativas para localizar QTLs} 

\posttitle{\par\end{flushleft}\vskip 0.5em} 

\preauthor{\begin{flushleft}\large \lineskip 0.5em \usefont{OT1}{phv}{b}{sl} \color{DarkRed}} 

\author{Filippo Biscarini, Stefano Biffani, Alessandra Stella } 

\postauthor{\footnotesize \usefont{OT1}{phv}{m}{sl} \color{Black} 
PTP (Parco Tecnologico Padano), Lodi, Italy 

\par\end{flushleft}\HorRule} 

\date{April 2015} 

\begin{document}

\maketitle 

\thispagestyle{fancy} 


\initial{S}\textbf{e presentan dos m\'{e}todos que se pueden utilizar como
  alternativas al GWAS para localizar QTLs en ganado. El primero se
  centra en el an\'{a}lisis de secuencias de homocigosis (“runs of
  homozygosity”, ROH) en grupos de animales (p.ej. casos y controles);
  el segundo es un m\'{e}todo de remuestreo basado en la frecuencia de inclusi\'{o}n de los SNP en modelos
predictivos. Las ROH se aplicaron a la identificaci\'{o}n de regiones
del genoma asociadas a trastornos reproductivos en vacas de raza Frisona, y el m\'{e}todo de remuestreo a la
detecci\'{o}n de portadores del haplotipo BH2 en vacas de raza Pardo
Alpina. Estos m\'{e}todos alternativos pueden complementar el GWAS est\'{a}ndar en la
localizaci\'{o}n de QTLs para caracteres de inter\'{e}s en el genoma de animales dom\'{e}sticos.}

[English abstract]
\initial{B}\textbf{eyond GWAS: alternatives to localize QTLs in farm
  animals. Two methods that could be used for QTL mapping as alternatives to standard GWAS are
presented. The first relies on the differential frequency of runs of homozygosity (ROH) in
groups of animals (e.g. cases and controls), while the second stems from resampling
techniques used for the prediction of carriers of a mutation, and is based on the frequency of
inclusion of polymorphisms (SNP) in the predictive model. ROH were applied to the detection
of reproductive diseases in Holstein-Friesian cattle, while resampling was applied to the
detection of carriers of the BH2 haplotype in Brown Swiss cattle. These alternative
approaches may complement GWAS analyses in localizing more accurately QTLs for traits of
interest in livestock.}


\section*{Introducci\'{o}n}

El descubrimiento de regiones del genoma asociadas a caracteres de inter\'{e}s zoot\'{e}cnico
(p.ej. resistencia a enfermedades) es una importante etapa preliminar para posibles
aplicaciones de la gen\'{o}mica a la cr\'{i}a de animales dom\'{e}sticos como la identificaci\'{o}n de
portadores de mutaciones favorables/desfavorables o la selecci\'{o}n asistida por marcadores.
Esto es ahora posible gracias a la gran cantidad de datos procedentes de las nuevas
t\'{e}cnicas de secuenciaci\'{o}n masiva (NGS: next generation sequencing) en forma p.ej. de
genotipados de alta densidad de SNPs y de secuencias de genoma completos. Los estudios
de asociaci\'{o}n de genoma completo (GWAS) se han convertido por lo tanto en una t\'{e}cnica
est\'{a}ndar para escanear el genoma en busca de polimorfismos asociados al fenotipo
analizado. Sin embargo, los m\'{e}todos de GWAS presentan algunas limitaciones: analizan
generalmente un SNP a la vez, son susceptibles de producir resultados espurios y, salvo en
caso de se\~{n}ales de asociaci\'{o}n muy patentes, pueden resultar de
difícil interpretaci\'{o}n (\cite{mccarthy2008genome})
.
En este art\'{i}culo se presentan dos alternativas al GWAS para localizar QTLs en ganado: el
an\'{a}lisis de secuencias de homocigosis (``runs of homozygosity'', ROH) y un m\'{et}odo de
remuestreo basado en la frecuencia de inclusi\'{o}n de los SNP en modelos predictivos.


\subsection*{Material y M\'{e}todos}

Para ilustrar las diferentes propiedades de los m\'{e}todos basados en GWAS, ROH y
remuestreo para localizar QTLs, se utilizaron los resultados de tres estudios previos con
bovinos lecheros de raza Frisona y Pardo Alpina. En particular: el GWAS se bas\'{o} en 17
vacas afectadas por el desplazamiento abomasal izquierdo y 47 controles
de raza Frisona (\cite{biffani2014adding}); el an\'{a}lisis de ROH en 163 vacas afectadas por trastornos reproductivos
y 295 controles de raza Frisona (\cite{biscarini2014applying}; y el m\'{e}todo de remuestreo en 513
portadores del haplotipo BH2 en BTA19 y 3132 no portadores de raza Pardo
Alpina (\cite{biffani2015predicting}). Todos los animales se genotiparon
con el chip bovino 50k de $54\,001$ SNPs.
\paragraph{GWAS}: El estudio de asociaci\'{o}n pangen\'{o}mico (GWAS) para el desplazamiento abomasal
izquierdo se llev\'{o} a cabo siguiendo el procedimiento GRAMMAR-GC
descrito por \cite{aulchenko2007genomewide}: primero, se ajust\'{o} un modelo de regresi\'{o}n log\'{i}stica para datos binarios
(caso-control) incluyendo los efectos sistem\'{a}ticos de granja, orden de parto y el efecto
polig\'{e}nico aleatorio, y posteriormente se utilizaron los residuos de este modelo en una
regresi\'{o}n lineal para estimar el efecto de cada SNP.
\paragraph{An\'{a}lisis de secuencias de homocigosis (ROH)}: Bajo la hip\'{o}tesis de que las
enfermedades complejas presentan una componente genética constituida por muchas
variantes recesivas distribuidas a lo largo del genoma, se analizaron las secuencias de
homocigosis (ROH) para identificar regiones asociadas a trastornos reproductivos en vacas
lecheras. Las ROH se definen como secuencias de genotipos homocigotos adyacentes, que
reflejan la transmisi\'{o}n de haplotipos id\'{e}nticos por parte de progenitores comunes. En lugar
de centrarse en un locus, las ROH consideran tambi\'{e}n al entorno, teniendo as\'{i} en cuenta el
cambio de frecuencia de SNP cercanos que est\'{e}n en desequilibrio de ligamiento con el
locus analizado. Se identificaron entonces las ROH presentes en la poblaci\'{o}n de vacas bajo
estudio a trav\'{e}s de la estimaci\'{o}n de la homocigosis de los SNP en ``ventanas deslizantes'' de
1000 kb. Se permitieron como m\'{a}ximo 5 genotipos faltantes y 1 genotipo heterocigoto para
considerar la secuencia como ROH. La frecuencia relativa de las ROH en casos y controles
indica regiones del genoma asociadas con el fenotipo.
\paragraph{Remuestreo de SNPs en modelos predictivos}: Para la identificaci\'{o}n de portadores del
haplotipo BH2 en vacas de raza Pardo Alpina se aplic\'{o} un procedimiento
basado en la reducci\'{o}n progresiva del n\'{u}mero de SNP incluidos en el modelo y en el an\'{a}lisis
discriminante lineal (LDA). Para cada umbral de SNP utilizados (2.5\%, 10\%, 15\%, 30\%, 50\%
y 100\% sobre el total de SNP) se seleccionaron aquellos m\'{a}s predictivos a trav\'{e}s del
m\'{e}todo de Best Subset Selection (BSS) y se utilizaron los SNP seleccionados para clasificar
los individuos en portadores o no del haplotipo. El procedimiento se repiti\'{o} 1000 veces
(validaci\'{o}n cruzada de 10 pliegues, 100 r\'{e}plicas) por cada umbral, y se obtuvo la gr\'{a}fica de
la frecuencia de inclusi\'{o}n de los SNP en el modelo predictivo en funci\'{o}n de su posici\'{o}n en
BTA19. De esta manera se pudo localizar la mutaci\'{o}n en el genoma.


\subsection*{Resultados y Discusi\'{o}n}

\begin{figure*}
\begin{center}
\includegraphics[width=0.95\textwidth]{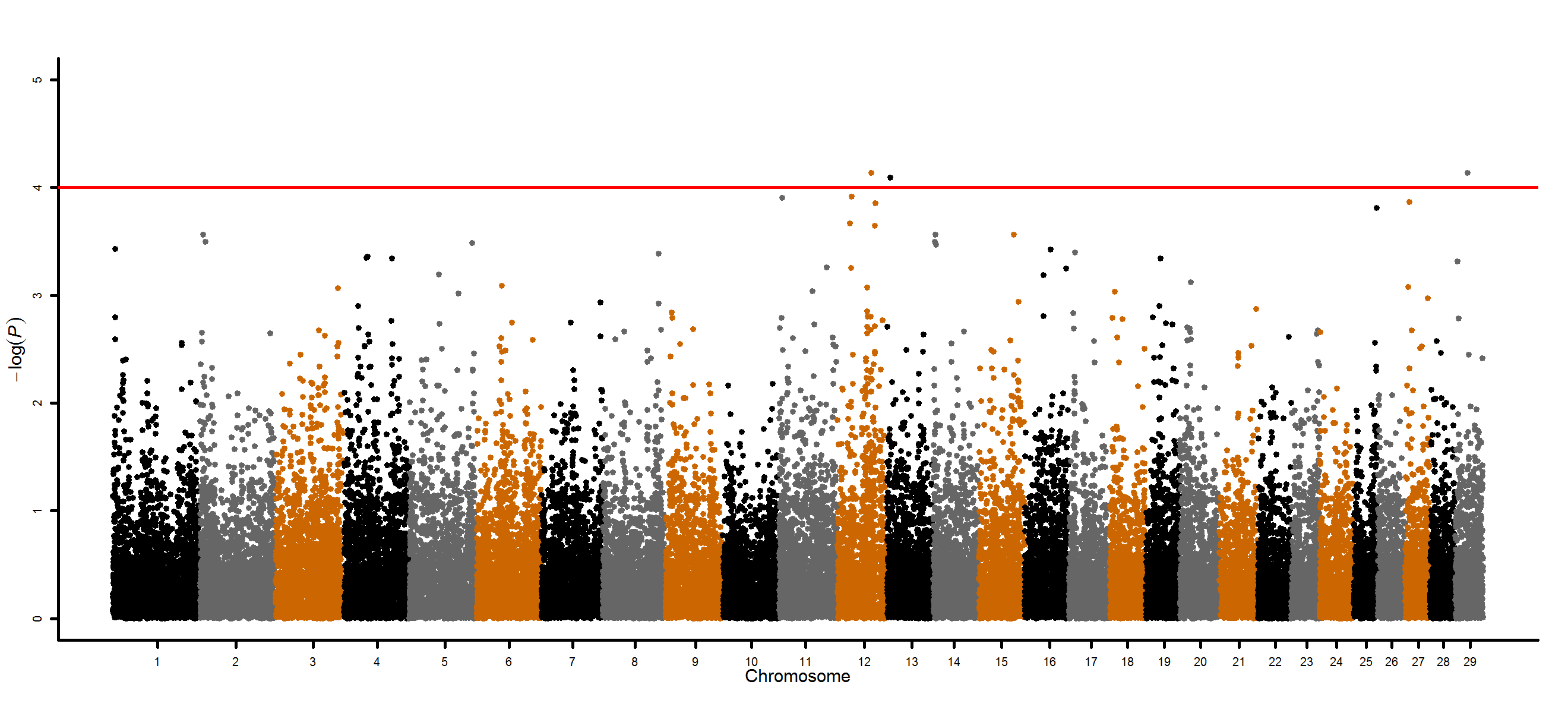}
\end{center}
\caption{Manhattan plot de los resultados ($–log(p\_value)$) del estudio de asociaci\'{o}n entre SNP y desplazamiento del
ab\'{o}maso en vacas de raza Frisona Holandesa. La l\'{i}nea roja indica
el umbral de significaci\'{o}n: $-log(0.0001) = 4$}
\label{fig:gwas}
\end{figure*}

En el estudio de asociaci\'{o}n se detectaron SNPs asociados con el
desplazamiento abomasal izquierdo en BTA12. La asociaci\'{o}n m\'{a}s
significativa se halla a una distancia de tan solo 20 kb del gen
\emph{SLITRK5}. Este gen pertenece a la familia g\'{e}nica \emph{SLITRK}, implicada en
procesos neurol\'{o}gicos que ya se demostraron asociados a la patog\'{e}nesis
del desplazamiento abomasal en vacuno. La Figura~\ref{fig:gwas} presenta el Manhattan
plot del an\'{a}lisis GWAS. Se puede apreciar la falta de se\~{n}ñales claras de
asociaci\'{o}n y la aparici\'{o}n de numerosas asociaciones, potencialmente
espurias, que resultan m\'{a}s o menos ``significativas'' y distribuidas a lo
largo del genoma.

\begin{figure*}
\begin{center}
\includegraphics[width=0.95\textwidth]{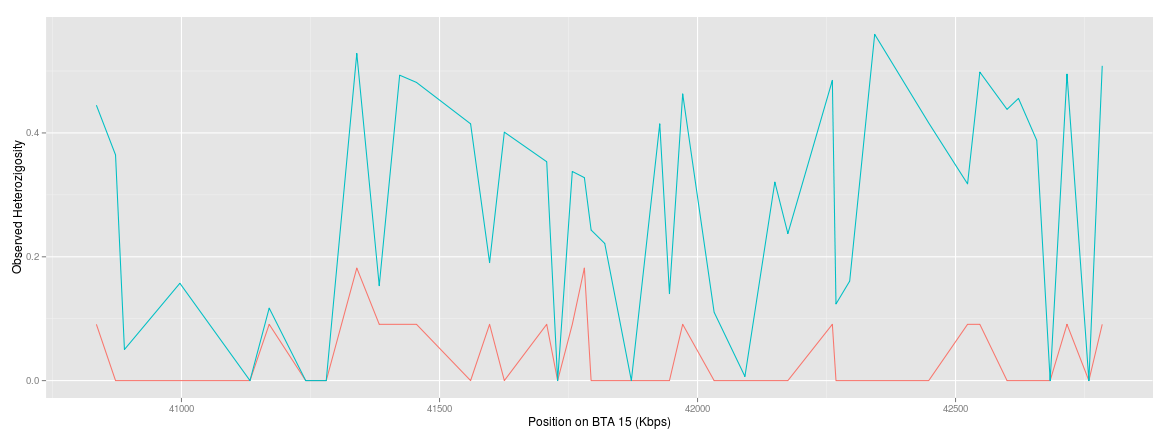}
\end{center}
\caption{Valores de homocigosis a lo largo del BTA15 en casos (rojo) y controles (azul) para enfermedades reproductivas.}
\label{fig:roh}
\end{figure*}

La interpretaci\'{o}n de los resultados puede ser
arbitraria, lo que ilustra algunas de las limitaciones del GWAS. El
an\'{a}lisis de las secuencias de homocigosis (ROH) permite ampliar la
mirada al entorno de cada SNP, llevando a interpretaciones m\'{a}s robustas
de los resultados: la aparici\'{o}n de un haplotipo homocigoto en casos pero
no en controles puede ser m\'{a}s informativa que un SNP asociado
individualmente. La Figura~\ref{fig:roh} ilustra el caso de trastornos reproductivos
en vacas lecheras en BTA15. El an\'{a}lisis de ROH es un m\'{e}todo falto de un
modelo estad\'{i}stico expl\'{i}cito, y esto puede causar problemas a la hora,
p.ej., de probar la significancia de la asociaci\'{o}n o de incluir efectos
sistem\'{a}ticos en el an\'{a}lisis.

Sin embargo, hay maneras de resolver estas dificultades. La significancia estad\'{i}stica se
puede testar comparando la homocigosis media (o la frecuencia de las
ROH) en casos y controles con un test-t, o aplicando el concepto
cl\'{i}nico de ``no inferioridad'' (\cite{d2003non}). Se realizan dos
an\'{a}lisis, ROH y GWAS, y se calcula el ``false discovery rate'' (FDR) para los dos: la hip\'{o}tesis
nula es que el FDR es mayor en ROH que en GWAS (ROH inferior al GWAS), la hip\'{o}tesis
alternativa es que los dos m\'{e}todos son equivalentes (mismo FDR medio). Los efectos
sistem\'{a}ticos se pueden incluir utilizando un modelo estad\'{i}stico para analizar las ROH
(\cite{pollott2012}); otra alternativa es estratificar el an\'{a}lisis de ROH seg\'{u}n las clases del efecto
sistem\'{a}tico (ej. machos/hembras), o aplicarlo a los residuos de un modelo lineal. M\'{a}s detalles
se encuentran en \cite{biscarini2014using,biscarini2014applying}.

\begin{figure*}
\begin{center}
\includegraphics[width=0.95\textwidth]{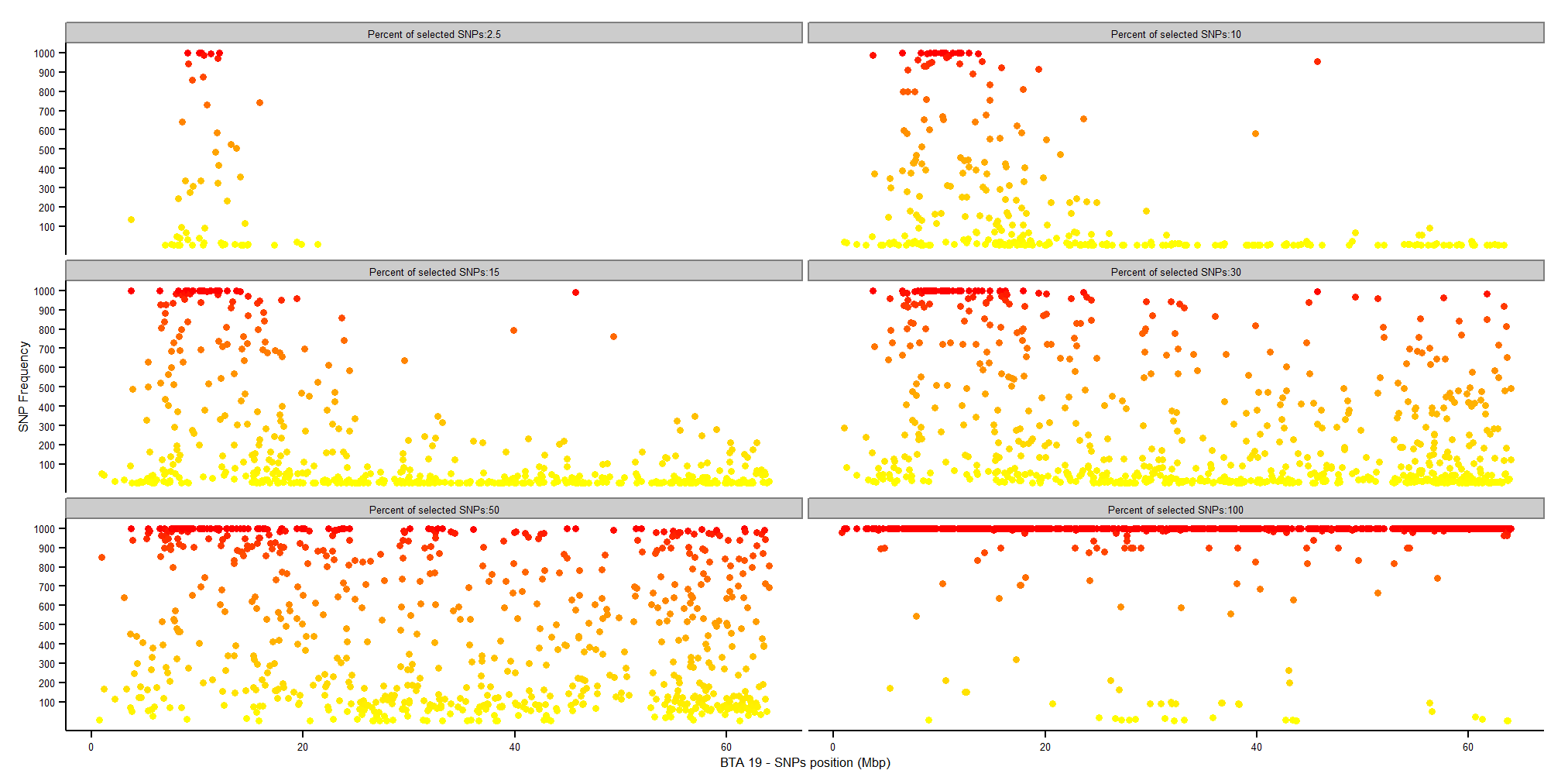}
\end{center}
\caption{Frecuencia de inclusi\'{o}n de los SNP en el modelo predictivo (en cada de las 1000 r\'{e}plicas por umbral) en funci\'{o}n
de la posici\'{o}n en BTA19.}
\label{fig:remuestreo}
\end{figure*}

Otra opci\'{o}n para localizar QTLs es utilizar los
resultados de t\'{e}cnicas de remuestreo. La Figura~\ref{fig:remuestreo} presenta la frecuencia de inclusi\'{o}n de los
SNP en el modelo predictivo para portadores del haplotipo BH2 en funci\'{o}n de la posici\'{o}n en
BTA19. Las t\'{e}cnicas de remuestreo permiten analizar la variabilidad alrededor del par\'{a}metro
estimado y se caracterizan por tener m\'{a}s poder estad\'{i}stico y localizar QTLs de manera m\'{a}s
robusta e independiente de los valores p y del ensayo de frecuencia a cada locus.
Los m\'{e}todos propuestos pueden ser utilizados para complementar el GWAS est\'{a}ndar en la
localizaci\'{o}n de QTLs para caracteres de inter\'{e}s en el genoma de animales
dom\'{e}sticos.



\bibliographystyle{apalike}
\bibliography{sample}






\end{document}